\documentclass[showpacs, secnumarabic, nobibnotes,  nofootinbib, aps,
prd]{revtex4}%
\usepackage{amsmath}
\usepackage{amssymb}
\usepackage{bm}
\usepackage{graphicx}
\usepackage{amsfonts}%
\providecommand{\U}[1]{\protect\rule{.1in}{.1in}}
\newtheorem{theorem}{Theorem}

\newtheorem{lemma}[theorem]{Lemma}

\begin{document}
\title{Spin structure of hadrons and minimum energy of bound systems}
\author{Petr Zavada}
\email{zavada@fzu.cz}
\affiliation{Institute of Physics of the Czech Academy of Sciences, Na Slovance 2, CZ-182
21 Prague 8, Czech Republic}

\begin{abstract}
The spin of a composite particle, like a nucleus or a hadron, is generated by
the composition of angular moments (consisting of spins and orbital angular
moments) of the constituents. The composition of two angular moments is done
by the standard way with the use of Clebsch-Gordan coefficients. However, if
there are more than two constituents, the composition must be done in a
hierarchical way, which admits more ways leading to the same resulting spin
state $\left\vert J,J_{z}\right\rangle $. Different composition patterns can
generate states with the same spin quantum numbers, but which may vary in the
contributions of different kinds of the constituents. We will discuss which
composition patterns could be preferred in the hadrons from the viewpoint of
minimal energy of the bound system. In this context, particular attention is
paid to the role of gluons or quark orbital angular momentum in the proton
spin.\medskip\ 

\end{abstract}

\maketitle

\section{Introduction}

Spin structure of hadrons that consist of quarks and gluons generating bond
among them is a difficult nonperturbative QCD problem. The complexity of this
task can be illustrated by a long history of the proton spin puzzle that began
with the EMC experiment \cite{emc}. Let's remind the main facts.

Before the well-known surprising result of the EMC experiment, it was expected
that the proton spin is simply defined by the sum of the spins of quarks
inside the proton. This expectation was disproved by measuring the structure
function $g_{1}(x)$, whose integral%
\begin{equation}
\Gamma_{1}=\int g_{1}(x)dx \label{ps}%
\end{equation}
allows us to evaluate the total contribution of the quark spins. It has been
shown that the spins of quarks represent only a small part of the proton spin.
This result was confirmed by further experiments that followed. At present,
the precise measurement of the Compass experiment \cite{compsig} implies the
spins of quarks represent only about $1/3$ of the proton spin. The small quark
contribution $\Delta\Sigma$ raised the question of how the proton spin is
actually generated? It is currently assumed that two additional contributions
may be relevant:

i) The total angular momentum of gluons can make a significant contribution.
If gluons can make up about half of the proton energy (mass), why would they
not significantly contribute to the spin? And if so, how much?

ii) Quark orbital angular momentum (OAM) can also significantly contribute to
the proton spin. The intrinsic motion of quarks localized inside the proton
can generate OAM, but how large?

The present study aims to discuss the task of AMs composition in the many-body
quantum mechanical systems and the implications related to these questions.
The main goal of this note is to show the argument why the gluon contribution
to the spin of most stable hadrons should be rather small (Sec.2). We also
remind our former arguments in favor of the important role of quark OAM
(Sec.3). \ A short discussion is presented in Sec.4.

\section{Interplay of the AMs inside the hadron}

Hadrons are the eigenstates $\Psi=|J,J_{z}\rangle$ of \textit{two} commuting
spin operators $\hat{J}^{2}=\hat{J}_{x}^{2}+\hat{J}_{y}^{2}+\hat{J}_{z}^{2}$
and $\hat{J}_{z}$ with the eigenvalues%
\begin{equation}
\hat{J}^{2}\Psi=J\left(  J+1\right)  \hbar^{2}\Psi,\quad\hat{J}_{z}\Psi
=J_{z}\hbar\Psi;\quad-J\leq J_{z}\leq J, \label{ps1}%
\end{equation}
where $J$ and $J_{z}$ are integers for mesons and half-integers for baryons.
The operators $\hat{J}_{x},\hat{J}_{y}$ and $\hat{J}_{z}$\ represent the spin
projections on three axes $x,y,z$, AM is a three-dimensional quantity. The
hadrons are composite particles consisting of quarks and gluons, so their spin
(total AM) is generated by the AM of their constituents, quarks and gluons.
The AM operators consist of the corresponding spin and OAM:%
\begin{equation}
\hat{\jmath}=\hat{s}+\hat{l}. \label{ps9}%
\end{equation}
The hadron rest frame is natural (initial) reference frame for their composition.

\subsection{Composition of the AMs}

\begin{figure}[t]
\centering\includegraphics[width=16cm]{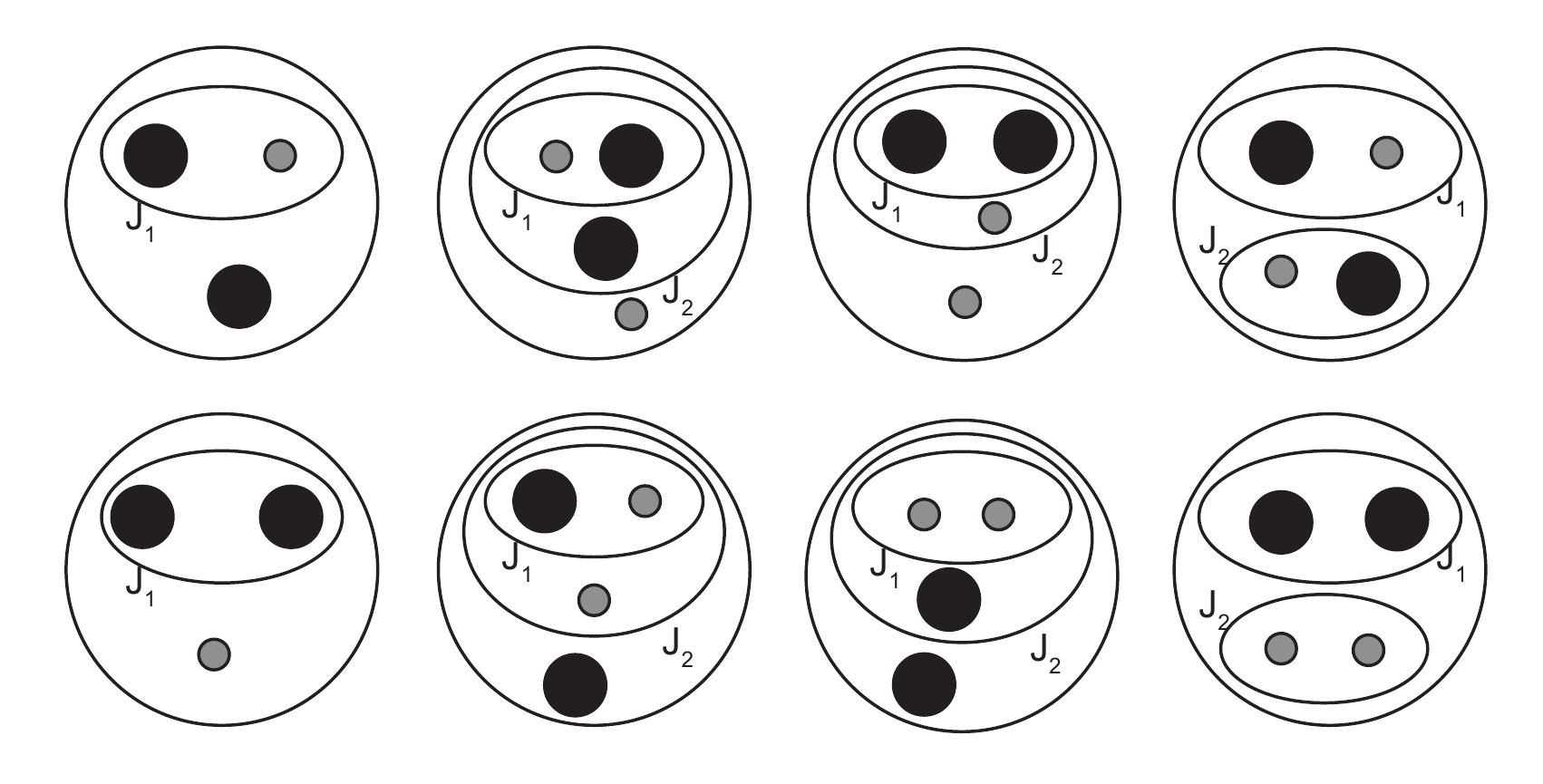}\caption{Examples of AM
composition patterns for meson. Greater black spots represent quarks, smaller
grey ones are gluons.}%
\label{figm}%
\end{figure}

\begin{figure}[t]
\centering\includegraphics[width=16cm]{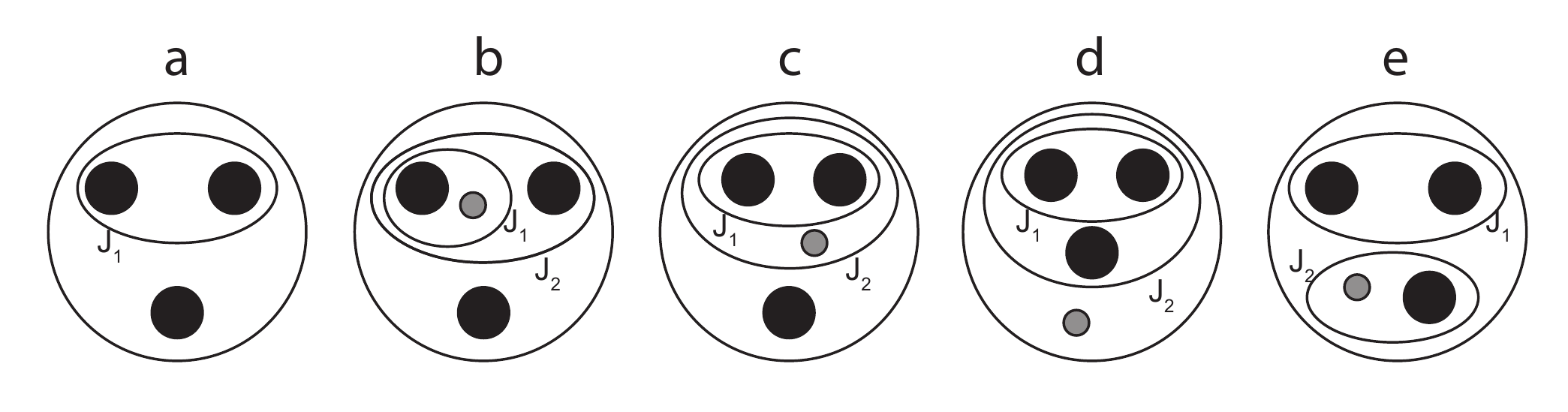}\caption{Examples of AM
composition patterns for baryon. The spots are quarks and gluons, like in the
previous picture.}%
\label{figh}%
\end{figure}

In quantum mechanics the binary AM composition is defined as%
\begin{align}
\left\vert (j_{1},j_{2})J,J_{z}\right\rangle  &  =\sum_{j_{z1}=-j_{1}}^{j_{1}%
}\sum_{j_{z2}=-j_{2}}^{j_{2}}\left\langle j_{1},j_{z1},j_{2},j_{z2}\left\vert
J,J_{z}\right.  \right\rangle \left\vert j_{1},j_{z1}\right\rangle \left\vert
j_{2},j_{z2}\right\rangle ;\label{rs9}\\
j_{z1}+j_{z2}  &  =J_{z},\qquad\qquad\qquad\left\vert j_{1}-j_{2}\right\vert
\leq J\leq j_{1}+j_{2}, \label{rs9a}%
\end{align}
where $\left\langle j_{1},j_{z1},j_{2},j_{z2}\left\vert J,J_{z}\right.
\right\rangle $ are Clebsch-Gordan coefficients, which are non-zero only if
the conditions (\ref{rs9a}) are satisfied. The symbols $j_{i},j_{zi}$ are
related to the total AM of the constituents (\ref{ps9}), $J,J_{z}$ are the
spin numbers of the hadron. The relation represents rule for AM composition of
$j_{1},j_{2}$ resulting in $J,$ symbolically $\left(  j_{1}\oplus
j_{2}\right)  _{J}$. If there are more then two AMs to compose, one must
repeat the binary composition to obtain the many-particle eigenstates of
resulting $J,J_{z}$%
\begin{equation}
\left\vert (j_{1},j_{2},...j_{n})_{c}J,J_{z}\right\rangle =\sum_{j_{z1}%
=-j_{1}}^{j_{1}}\sum_{j_{z2}=-j_{2}}^{j_{2}}...\sum_{j_{zn}=-j_{n}}^{j_{n}%
}c_{j}\left\vert j_{1},j_{z1}\right\rangle \left\vert j_{2},j_{z2}%
\right\rangle ...\left\vert j_{n},j_{zn}\right\rangle , \label{rs10}%
\end{equation}
where the coefficients $c_{j}$ consist of the Clebsch-Gordan coefficients%
\begin{equation}
c_{j}=\left\langle j_{1},j_{z1},j_{2},j_{z2}\left\vert J_{3},J_{3z}\right.
\right\rangle \left\langle J_{3},J_{z3},j_{3},j_{z3}\left\vert J_{4}%
,J_{z4}\right.  \right\rangle ...\left\langle J_{n},J_{zn},j_{n}%
,j_{zn}\left\vert J,J_{z}\right.  \right\rangle . \label{rs10b}%
\end{equation}
However the set $j_{1},j_{2},..j_{n}$ does not define the resulting state $J$
unambiguously, this state depends also on the intermediate values $J_{1}%
,J_{2}...$, and the order of composition. The subscript $c$ in l.h.s. of
(\ref{rs10}) denotes a definite pattern of composition. In Fig. \ref{figm} we
show symbolic examples of a few composition patterns for two quarks and one or
two gluons. Similar examples for three quarks and one gluon follow in Fig.
\ref{figh}. The patterns in the figures can be labeled as%
\begin{equation}
c=(((j_{1}\oplus j_{2})_{J_{1}}\oplus j_{3})_{J_{2}}\oplus k_{1})_{J}
\label{rs10c}%
\end{equation}
for Fig. \ref{figh}d and similarly for others. The symbol (\ref{rs10c})
represents three consecutive steps of composition (\ref{rs9}) giving the
states%
\begin{equation}
\left\vert (j_{1},j_{2})J_{1},J_{1z}\right\rangle ,\qquad\left\vert
(J_{1},j_{3})J_{2},J_{2z}\right\rangle ,\qquad\left\vert (J_{2},k)J,J_{z}%
\right\rangle , \label{ps3}%
\end{equation}
where $J_{i}$ represent intermediate AMs:%
\begin{equation}
j_{1}\oplus j_{2}=J_{1},\qquad J_{1}\oplus j_{3}=J_{2},\qquad J_{2}\oplus k=J.
\label{10e}%
\end{equation}
A priory various patterns for a fixed\ $J$ can be allowed, but their number
increases with the number of the constituents in the composition very fast.
The real state can be some superposition of the Fock states defined by these
patterns. At the same time in the real scenario of interacting particles, one
can expect their probabilities will vary, but which patterns will be
preferred? And how can the states with the same resulting spin differ?

\subsection{Bound systems}

We work with the many particle stationary states that are controlled by the
unperturbative QCD, which does not give much hope for a simple solution. On
the other hand, even without knowledge of the solution of complicated
equations it is known that stable or quasi-stable systems are always
associated with some (rest frame) energy minimum. So we will discuss which
type of patterns can prefer lower energy minimum.

Let us recall the role of AM in a few well-known examples of bound, stable or
quasi-stable systems:

\textit{a)} In atoms higher OAM of electrons correlate with higher energy
levels. At the same time, the electrons in noble gases having fully occupied
energy levels with total AM equal zero are the most stable atoms. Noble gases
have the largest ionization potential for each period.

\textit{b)} Even-even nuclei have always total spin $J=0$ and are known to be
more stable. Some good examples are $_{2}^{4}$He, $_{6}^{12}$C and $_{8}^{16}$O.

\textit{c)} The excited states of mesons and nucleons are associated with
higher masses (energy) together with higher spins. For example $\rho
(770)^{0\pm}$ mesons are excited states of pions, resonances $\Delta
(1232)^{0+}$ are excited neutron and proton. The empirical almost linear
dependencies $M^{2}$ on $J$ for the families of meson and baryon resonances
\cite{regge} are known as the Regge trajectories.

The systems \textit{b),c)} are controlled by the QCD. The examples suggest the
correlation:%
\[
energy\ (mass)\sim total\ AM\ (spin),\quad minimum\ energy\sim minimum\ AM
\]
Obviously:

$\bullet$ Minimum energy is a manifestation of the bond

$\bullet$ Bond generates a tendency to minimum AM:

Presence of gluons (color field) mediating bond among quarks generates lower
energy, which generates minimum AM of the corresponding state (or segment).
The minimum energy of gluons is not influenced by the Pauli principle. So, we
shall try to work with the rule:

\textit{Presence of the gluons is preferred in the segments of }%
$J=0$\textit{:}%
\begin{equation}
(...(...\oplus k)...)_{J=0}, \label{ps5}%
\end{equation}
where $k$ is the gluon AM. It is important that any composition pattern for
particles of arbitrary AMs $j_{1},j_{2},...j_{n\text{ }}$ generating the state
$J=0$, gives also the \textit{condition for all} $j_{zi}:$
\begin{equation}
\left\langle j_{z1}\right\rangle =\left\langle j_{z2}\right\rangle
=....=\left\langle j_{zn}\right\rangle =0. \label{ps6}%
\end{equation}
This rule together with the relation%
\begin{equation}
\left\langle j_{z1}\right\rangle +\left\langle j_{z2}\right\rangle
+....+\left\langle j_{zn}\right\rangle =J_{z}, \label{ps7}%
\end{equation}
are proved in Appendix. We will discuss consequences of (\ref{ps5}) for
hadrons with minimum spin: $J=0$ (mesons) and $J=1/2$ (baryons). Obviously,
obtained condition (\ref{ps6}) is stronger than%
\begin{equation}
\left\langle j_{z1}\right\rangle +\left\langle j_{z2}\right\rangle
+....+\left\langle j_{zn}\right\rangle =0. \label{ps6a}%
\end{equation}
To simplify\ the discussion, we do not strictly distinguish quarks, antiquarks
and their flavors, only their AMs are important.

\subsection{Mesons of spin $J=0$}

We assume arbitrary patterns involving $2n$ quarks and any number of gluons,
which generate the state $J=J_{z}=0$. Examples of such systems are mesons
$\pi,\eta,K,D,$ etc. The rule of preference (\ref{ps5}) is satisfied
automatically, then the rule (\ref{ps6}) means that%
\begin{equation}
\left\langle j_{z}^{a}\right\rangle =0, \label{ps8}%
\end{equation}
where $a$ denotes constituents: quarks and gluons. It means that average AMs
of gluons and quarks of any flavor are zero in the scalar or pseudoscalar mesons.

\subsection{Baryons of spin $J=1/2$}

Now we assume arbitrary patterns with $3+2n$ quarks and any number of gluons,
which generate states of minimum spin content, $J=J_{z}=1/2$. The examples in
Fig. \ref{figh} are completed by Tab. \ref{tbl1}.

\begin{table}[ptb]
\caption{Examples of composition patterns}%
\label{tbl1}
\centering%
\begin{tabular}
[c]{|l|c|c|}\hline
& Composition pattern & $\left\langle j_{z1}\right\rangle ,\left\langle
j_{z2}\right\rangle ,\left\langle j_{z3}\right\rangle ,\left\langle
k_{z}\right\rangle $\\\hline
a & \multicolumn{1}{|l|}{$\left(  \left(  j_{1}\oplus j_{2}\right)  _{0}\oplus
j_{3}\right)  _{1/2}$} &
\multicolumn{1}{|l|}{$\ \ \ 0,\ \ \ \ \ 0,\ \ \ \ \frac{1}{2},\ \ \ 0$%
}\\\hline
b$_{1}$ & \multicolumn{1}{|l|}{$\left(  \left(  \left(  j_{1}\oplus k\right)
_{1/2}\oplus j_{2}\right)  _{0}\oplus j_{3}\right)  _{1/2}$} &
\multicolumn{1}{|l|}{$\ \ \ 0,\ \ \ \ \ 0,\ \ \ \ \frac{1}{2},\ \ \ 0$%
}\\\hline
b$_{2}$ & \multicolumn{1}{|l|}{$\left(  \left(  \left(  j_{1}\oplus k\right)
_{1/2}\oplus j_{2}\right)  _{1}\oplus j_{3}\right)  _{1/2}$} &
\multicolumn{1}{|l|}{$\frac{-1}{9},\ \ \ \ \frac{1}{3},\ \ \frac{-1}%
{6},\ \ \frac{4}{9}$}\\\hline
c & \multicolumn{1}{|l|}{$\left(  \left(  \left(  j_{1}\oplus j_{2}\right)
_{1}\oplus k\right)  _{0}\oplus j_{3}\right)  _{1/2}$} &
\multicolumn{1}{|l|}{$\ \ \ 0,\ \ \ \ \ 0,\ \ \ \ \frac{1}{2},\ \ 0$}\\\hline
d$_{1}$ & \multicolumn{1}{|l|}{$\left(  \left(  \left(  j_{1}\oplus
j_{2}\right)  _{0}\oplus j_{3}\right)  _{1/2}\oplus k\right)  _{1/2}$} &
\multicolumn{1}{|l|}{$\ \ \ 0,\ \ \ \ \ 0,\ \ \frac{-1}{6},\ \frac{2}{3}$%
}\\\hline
d$_{2}$ & \multicolumn{1}{|l|}{$\left(  \left(  \left(  j_{1}\oplus
j_{2}\right)  _{1}\oplus j_{3}\right)  _{1/2}\oplus k\right)  _{1/2}$} &
\multicolumn{1}{|l|}{$\frac{-1}{9},\ \ \frac{-1}{9},\ \ \ \frac{1}{18}%
,\ \frac{2}{3}$}\\\hline
e$_{1}$ & \multicolumn{1}{|l|}{$\left(  \left(  j_{1}\oplus j_{2}\right)
_{0}\oplus\left(  j_{3}\oplus k\right)  _{1/2}\right)  _{1/2}$} &
\multicolumn{1}{|l|}{$\ \ \ 0,\ \ \ \ \ 0,\ \ \frac{-1}{6},\ \frac{2}{3}$%
}\\\hline
e$_{2}$ & \multicolumn{1}{|l|}{$\left(  \left(  j_{1}\oplus j_{2}\right)
_{1}\oplus\left(  j_{3}\oplus k\right)  _{1/2}\right)  _{1/2}$} &
\multicolumn{1}{|l|}{$\ \ \frac{1}{3},\ \ \ \frac{1}{3},\ \ \frac{1}%
{18},\ \frac{-2}{9}$}\\\hline
\end{tabular}
\end{table}All patterns in figure generate composite fermion of the same spin,
but in general the states are different. The patterns differ in\ the right
column of the table, where the quark and gluon contributions vary, but
satisfy
\begin{equation}
\left\langle j_{z1}\right\rangle +\left\langle j_{z2}\right\rangle
+\left\langle j_{z3}\right\rangle +\left\langle k_{z}\right\rangle =\frac
{1}{2}, \label{ps2}%
\end{equation}
which is a special case of (\ref{ps7}).

a) A toy model: scenario $3q+gluons$

Obviously the preferred segments (\ref{ps5}) are present only in the patterns
\textbf{b}$_{1}$ and \textbf{c} in table and in corresponding panels of Fig.
\ref{figh}. Here the gluon is shared by two quarks within an intermediate
state with $J=0$. Corresponding baryon state is generated by a superposition
of both patterns with the permutations of the three quarks. In a more general
version one can consider any number of gluons shared by two the quarks within
the state $J=0$. For example, there can be some correspondence between this
scenario and the known quark-diquark model \cite{q2q}. The rule of preference
(\ref{ps5}) suppresses the contribution of gluon AM.

b) General scenario $3q+nq\bar{q}+gluons$

The numbers of sea quarks and gluons are not fixed and baryon state is
represented by a corresponding superposition of the Fock states $J=J_{z}=1/2,$
which are constrained by the rule of preference (\ref{ps5}). Preferred
segments in patterns \textbf{b}$_{1}$ and \textbf{c} in the table are just
simple examples. In general scenario, preferred segments can create more
complex quark-gluon states of spin $J=0$, for which the condition (\ref{ps6})
is satisfied. It follows, that average quark and gluon AM contributions are
zero in these segments. Since we assume the gluons sit only in preferred
segments (\ref{ps5}), the total gluon AM contribution must be suppressed in
the resulting Fock states. We talk about the total gluon AM and avoid the
controversies in its splitting to the spin and orbital part, as analyzed
thoroughly in \cite{Leader:2013jra}.

\subsection{Connection with QCD}

In general, the structure of hadrons results from QCD, which is represented by
a complex system of differential equations. A prerequisite for the solution of
the system is the specification of boundary conditions defining the hadron. In
addition to the composition of the bound system (a type of meson, baryon or
nucleus, including its spin) a condition of minimum energy in the hadron rest
frame is required. As we have suggested, minimum energy can be correlated with
the particular AM composition patterns. In this sense, our discussion is
related to the QCD boundary conditions.

\section{Reason for quark OAM}

We have studied the role of the quark OAM in the covariant quark-parton model
\cite{Zavada:2007ww,Zavada:2013ola,Zavada:2015gaa,
Zavada:2002uz,Zavada:2001bq}, the essence is as follows. For Dirac particle,
it is known that in relativistic case the spin $\left(  s,s_{z}\right)  $ and
OAM $\left(  l,l_{z}\right)  $ are not decoupled (separately conserved), but
only quantum numbers $j,j_{z}$ corresponding to the total angular momentum
(AM) are conserved ($j,j_{z}$ are the good quantum numbers). However, one can
always calculate the mean values $\left\langle s_{z}\right\rangle $ and
$\left\langle l_{z}\right\rangle ,$ and it holds
\begin{equation}
j_{z}=\left\langle s_{z}\right\rangle +\left\langle l_{z}\right\rangle .
\label{psa}%
\end{equation}
We have studied this issue in the representation of spinor spherical
harmonics, which allowed us to explore this relativistic spin-orbit interplay
more explicitly. From general relations%
\begin{equation}
\left\langle s_{z}\right\rangle _{j,j_{z}}=\frac{1+\left(  2j+1\right)  \mu
}{4j\left(  j+1\right)  }j_{z},\qquad\left\langle l_{z}\right\rangle
_{j,j_{z}}=\left(  1-\frac{1+\left(  2j+1\right)  \mu}{4j\left(  j+1\right)
}\right)  j_{z} \label{psa1}%
\end{equation}
we obtained for particle of spin $1/2$ in the state $j=j_{z}=1/2$%

\begin{equation}
\left\langle s_{z}\right\rangle =\frac{1+2\mu}{6},\qquad\left\langle
l_{z}\right\rangle =\frac{1-\mu}{3};\qquad\mu=\frac{m}{\epsilon}, \label{psb}%
\end{equation}
where $m$\ and \ $\epsilon$\ \ are mass and energy of the particle. So, for
non-relativistic $\left(  \mu\rightarrow1\right)  $ case we have
\begin{equation}
\left\langle s_{z}\right\rangle =j_{z}=1/2,\quad\left\langle l_{z}%
\right\rangle =0; \label{psc1}%
\end{equation}
and for relativistic one $\left(  \mu\rightarrow0\right)  $%
\begin{equation}
\left\langle s_{z}\right\rangle =1/6,\quad\left\langle l_{z}\right\rangle
=1/3,\quad\left\langle s_{z}\right\rangle +\left\langle l_{z}\right\rangle
=j_{z}=1/2. \label{pscr}%
\end{equation}
The last relation represents a kinematical effect in relativistic quantum
mechanics. This effect is exactly reproduced in the covariant quark-parton
model, where the effective value of $\mu$\ is a free parameter. Other
representations of the relativistic suppression of the quark spin are
discussed in the papers \cite{Zhang:2012sta, Ma:2001ui, Qing:1998at,
Liang:1996je,Ma:1992sj, Ma:1991xq}. The last relation, corresponding to
$\mu_{eff}\rightarrow0,$ represents the scenario of massless quarks. The quark
spin contribution to the proton spin is $\Sigma\approx1/3$ and the missing
part is balanced by the quark OAM. It means a very good agreement with the
data \cite{compsig} even without a gluon contribution. At the same time, the
same experiment \cite{compglu} suggests gluon contribution rather small. In
general scenario $\mu_{eff}>0$, then (\ref{psb}) implies that contribution of
the quark OAM should be less, so some gluon contribution is needed to explain
the data on $\Sigma$. Such a prediction could be consistent with the data
\cite{star}.

\section{Discussion and conclusion}

In our approach, the following conditions are necessary:

1) Also in the covariant parton model, we assume that in deep inelastic
scattering the quarks can be considered quasi-free. If this assumption is met
in the infinite momentum frame, it is also met in other reference systems -
including the nucleon rest frame \cite{Zavada:2013ola}.

2) Rest frame of the composite system (nucleon) with 3D rotation invariance is
necessary condition for quantum mechanical composition of spins and orbital
moments of the constituents, with the use of Clebsch-Gordan coefficients. At
the same time, for bound systems, the rest frame is necessary for the
expression of the minimum-energy condition.

The second condition allows us to use the representation of spinor spherical
harmonics, which implies the spin-orbit interplay is controlled by the ratio
$m/\epsilon$. Obviously, the 3D rotational invariance of the rest frame is not
fulfilled for the current light-cone formalism. That is why a similar
spin-orbit constraint is missing. In the covariant model, the assumptions
formulated in the rest frame generate the shape and properties of (invariant)
structure functions.

In this report we have studied the AM composition in many (quasi-free)
particle systems. We focused on hadrons and in particular on the proton. We
suggested argument, why the AM contribution of gluons to the proton spin
should be rather small. At the same time, we have discussed the important role
of the quark OAM following from kinematical effect of relativistic quantum
mechanics, which is simply reproduced in the covariant quark-parton model, but
not in the light-cone formalism.

\begin{acknowledgments}
This work was supported by the project LTT17018 of the MEYS (Czech Republic).
I am grateful to Peter Filip and Oleg Teryaev for useful discussions and
valuable comments.
\end{acknowledgments}

\appendix\textbf{ }

\section{\textbf{Proof of relations }(\ref{ps2}), (\ref{ps6})}

\label{appe}

\begin{lemma}
Any composition pattern for particles of arbitrary integer or half-integer
$j_{1},j_{2},...j_{n\text{ }}$ that generates state $J=0$, implies the
condition for all $j_{zi}:$
\begin{equation}
\left\langle j_{z1}\right\rangle =\left\langle j_{z2}\right\rangle
=....=\left\langle j_{zn}\right\rangle =0. \label{a1}%
\end{equation}

\end{lemma}

\textbf{Proof:} \ For the state%
\begin{equation}
\left\vert (j_{1},j_{2},...j_{n})_{c}J,J_{z}\right\rangle =\sum_{j_{z1}%
=-j_{1}}^{j_{1}}\sum_{j_{z2}=-j_{2}}^{j_{2}}...\sum_{j_{zn}=-j_{n}}^{j_{n}%
}c_{j}\left\vert j_{1},j_{z1}\right\rangle \left\vert j_{2},j_{z2}%
\right\rangle ...\left\vert j_{n},j_{zn}\right\rangle , \label{aa1}%
\end{equation}
where the coefficients $c_{j}$ consist of the Clebsch-Gordan coefficients%
\begin{equation}
c_{j}=\left\langle j_{1},j_{z1},j_{2},j_{z2}\left\vert J_{3},J_{3z}\right.
\right\rangle \left\langle J_{3},J_{z3},j_{3},j_{z3}\left\vert J_{4}%
,J_{z4}\right.  \right\rangle ...\left\langle J_{n},J_{zn},j_{n}%
,j_{zn}\left\vert J,J_{z}\right.  \right\rangle , \label{aa2}%
\end{equation}
we calculate average $j_{z1}$:
\begin{equation}
\left\langle j_{z1}\right\rangle =\sum_{j_{z1}=-j_{1}}^{j_{1}}j_{z1}%
\sum_{j_{z2}=-j_{2}}^{j_{2}}...\sum_{j_{zn}=-j_{n}}^{j_{n}}c_{j}c_{j}^{\ast
}=\sum_{j_{z1}=-j_{1}}^{j_{1}}j_{z1}w\left(  j_{z1}\right)  . \label{a3}%
\end{equation}
The Clebch-Gordan coefficients are real and it holds:
\begin{equation}
\left\langle j_{k},j_{zk},j_{l},j_{zl}\left\vert J_{i},J_{iz}\right.
\right\rangle =\left(  -1\right)  ^{J_{i}-j_{k}-j_{l}}\left\langle
j_{k},-j_{zk},j_{l},-j_{zl}\left\vert J_{i},-J_{iz}\right.  \right\rangle .
\label{a4}%
\end{equation}
For $J=J_{z}=0$ the term%
\begin{equation}
w\left(  j_{z1}\right)  =\sum_{j_{z2}=-j_{2}}^{j_{2}}...\sum_{j_{zn}=-j_{n}%
}^{j_{n}}c_{j}c_{j}^{\ast}>0 \label{a5}%
\end{equation}
is\ in the sum (\ref{a3})\ always accompanied by the term $w\left(
-j_{z1}\right)  $ and
\begin{equation}
w\left(  j_{z1}\right)  =w\left(  -j_{z1}\right)  , \label{a6}%
\end{equation}
which implies $\left\langle j_{z1}\right\rangle =0$. Similarly for others in
(\ref{a1}).

\textit{ii) Relation (\ref{ps7})}

Eq. (\ref{rs10}) implies%

\begin{align}
\left\langle j_{z1}\right\rangle +\left\langle j_{z2}\right\rangle
+...\left\langle j_{zn}\right\rangle  &  =\left\langle j_{z1}+j_{z2}%
+...j_{zn}\right\rangle \nonumber\\
&  =\sum_{j_{z1}=-j_{1}}^{j_{1}}\sum_{j_{z2}=-j_{2}}^{j_{2}}...\sum
_{j_{zn}=-j_{n}}^{j_{n}}c_{j}c_{j}^{\ast}\left(  j_{z1}+j_{z2}+...j_{zn}%
\right) \nonumber\\
&  =\sum_{j_{z1}=-j_{1}}^{j_{1}}\sum_{j_{z2}=-j_{2}}^{j_{2}}...\sum
_{j_{zn}=-j_{n}}^{j_{n}}c_{j}c_{j}^{\ast}J_{z}=J_{z}. \label{a2}%
\end{align}
Relation (\ref{ps2}) is a special case of this equation.


\begin{thebibliography}{99}                                                                                               %


\bibitem {emc}J. Ashman et al. [EMC Collaboration], Nucl. Phys. B
\textbf{238}, 1 (1990); Phys. Lett. B \textbf{206}, 364 (1988).

\bibitem {compsig}V. Y. .Alexakhin et al. [COMPASS Collaboration], Phys. Lett.
B \textbf{647}, 8 (2007).

\bibitem {compglu}C. Adolph et al. [COMPASS Collaboration], Phys. Lett. B
\textbf{718}, 922 (2013).

\bibitem {star}L. Adamczyk et al. [STAR Collaboration], Phys. Rev. Lett.
\textbf{115}, no. 9, 092002 (2015).

\bibitem {regge}Alfred Tang and John W. Norbury, Phys. Rev. D \textbf{62}, 016006(2000).

\bibitem {q2q}D. B. Lichtenberg and L. J. Tassie, Phys. Rev. \textbf{155}, 1601(1967).

\bibitem {Leader:2013jra}E.~Leader and C.~Lorc\'{e},
Phys.\ Rept.\ \textbf{541}, no. 3, 163 (2014).

\bibitem {Zavada:2015gaa}P.~Zavada,
Phys.\ Lett.\ B \textbf{751}, 525 (2015).

\bibitem {Zavada:2013ola}P.~Zavada,
Phys.\ Rev.\ D \textbf{89}, no. 1, 014012 (2014).

\bibitem {Zavada:2007ww}P.~Zavada,
Eur.\ Phys.\ J.\ C \textbf{52}, 121 (2007).

\bibitem {Zavada:2002uz}P.~Zavada,
Phys.\ Rev.\ D \textbf{67}, 014019 (2003).

\bibitem {Zavada:2001bq}P.~Zavada,
Phys.\ Rev.\ D \textbf{65}, 054040 (2002).

\bibitem {Zhang:2012sta}X.~Zhang and B.~Q.~Ma,
Phys.\ Rev.\ D \textbf{85}, 114048 (2012).

\bibitem {Ma:2001ui}B.~Q.~Ma, I.~Schmidt and J.~J.~Yang,
Eur.\ Phys.\ J.\ A \textbf{12}, 353 (2001).

\bibitem {Qing:1998at}D.~Qing, X.~S.~Chen and F.~Wang,
Phys.\ Rev.\ D \textbf{58}, 114032 (1998).

\bibitem {Liang:1996je}Z.~T.~Liang and R.~Rittel,
Mod.\ Phys.\ Lett.\ A \textbf{12}, 827 (1997).

\bibitem {Ma:1992sj}B.~Q.~Ma,
Z.\ Phys.\ C \textbf{58}, 479 (1993).

\bibitem {Ma:1991xq}B.~Q.~Ma,
J.\ Phys.\ G \textbf{17}, L53 (1991) doi:10.1088/0954-3899/17/5/001
\end{thebibliography}
\end{document}